%% file: main.tex
\documentclass[11pt]{article}

\usepackage{amsthm}
\usepackage{graphicx} % support the \includegraphics command and options
\usepackage{array} % for better arrays (eg matrices) in maths

\usepackage{amsmath, amssymb, amsfonts, verbatim}
\usepackage{hyphenat,epsfig,subfigure,multirow}
\usepackage{algorithm}
\usepackage{algorithmicx}
\usepackage{algpseudocode}

\usepackage[usenames,dvipsnames]{xcolor}

\usepackage{tcolorbox}
\tcbuselibrary{skins,breakable}
\tcbset{enhanced jigsaw}

\usepackage[compact]{titlesec}

\definecolor{DarkRed}{rgb}{0.5,0.1,0.1}
\definecolor{DarkBlue}{rgb}{0.1,0.1,0.5}

\usepackage{hyperref}
\hypersetup{
colorlinks=true,
pdfnewwindow=true,
citecolor=ForestGreen,
linkcolor=DarkRed,
filecolor=DarkRed,
urlcolor=DarkBlue
}

\usepackage{bm}
\usepackage{url}
\usepackage{xspace}
\usepackage[mathscr]{euscript}

\usepackage{mdframed}

\usepackage{cite}
\usepackage{enumitem}

\usepackage[margin=1in]{geometry}

\newtheorem{theorem}{Theorem}
\newtheorem{lemma}{Lemma}[section]

\newtheorem{cor}[theorem]{Corollary}
\newtheorem{claim}[lemma]{Claim}

\newtheorem{defn}{Definition}
\newtheorem{hypo}{Hypothesis}

\newtheorem{problem}{Problem}

\newtheorem{obs}[lemma]{Observation}

\newtheorem*{claim*}{Claim}
\newtheorem*{proposition*}{Proposition}
\newtheorem*{lemma*}{Lemma}
\newtheorem*{problem*}{Problem}

\newtheorem{mdresult}[theorem]{Theorem}

\newtheorem{mdinvariant}{Invariant}

\newcommand{\Acknowledgement}[1]{\ \\{\bf Acknowledgement:} #1}

\allowdisplaybreaks

\renewcommand{\qed}{\nobreak \ifvmode \relax \else
      \ifdim\lastskip<1.5em \hskip-\lastskip
      \hskip1.5em plus0em minus0.5em \fi \nobreak
      \vrule height0.75em width0.5em depth0.25em\fi}

\input{macros}

\title{Improved Algorithms for Fully Dynamic Maximal Independent Set}

\author{Yuhao Du\thanks{{\small{\tt duyh15@mails.tsinghua.edu.cn}}} \\ IIIS, Tsinghua University
\and Hengjie Zhang\thanks{{\small{\tt zhanghj15@mails.tsinghua.edu.cn}}} \\ IIIS, Tsinghua University
}

\date{}

\begin{document}
\maketitle

\thispagestyle{empty}
\input{abstract}
%\clearpage
\setcounter{page}{0}

\newpage

\input{introduction}

\input{determinstic}

\input{random}

\input{hardness}

\Acknowledgement{We thank Seth Pettie for helpful discussions.}

\bibliographystyle{abbrv}
\bibliography{ref}

\end{document}

%% file: macros.tex
\newcommand{\toShrink}{-.20cm}
\newcommand{\toShrinkEnu}{-.2cm}

\DeclareMathOperator*{\Prob}{\ensuremath{\textnormal{Pr}}}
\renewcommand{\Pr}{\Prob}

\newenvironment{tbox}{\begin{tcolorbox}[
		enlarge top by=5pt,
		enlarge bottom by=5pt,
		 breakable,
		 boxsep=0pt,
                  left=4pt,
                  right=4pt,
                  top=10pt,
                  arc=0pt,
                  boxrule=1pt,toprule=1pt,
                  colback=white
                  ]%%
	}
{\end{tcolorbox}}

%% file: abstract.tex
\begin{abstract}
Maintaining maximal independent set in dynamic graph is a fundamental open problem in graph theory and the first sublinear time deterministic algorithm was came up by Assadi, Onak, Schieber and Solomon(STOC'18), which achieves $O(m^{3/4})$ amortized update time. We have two main contributions in this paper. We present a new simple deterministic algorithm with $O(m^{2/3}\sqrt{\log m})$ amortized update time, which improves the previous best result. And we also present the first randomized algorithm with expected $O(\sqrt{m}\log^{1.5}m)$ amortized time against an oblivious adversary. %For low arboricity graphs (e.g., planar graphs and graphs excluding fixed minors), we show another deterministic algorithm which runs in $O(\sqrt{m\log m})$ amortized time, which is also the first algorithm for fully dynamic MIS respect to the arboricity.
\end{abstract}

%% file: introduction.tex
\section{Introduction}

The model of dynamic graph is very common in real life like the connection between two servers break down by accident and be restored some time later. And it is of great interest to theoretical computer science field. We need to maintain a solution subject to some constraints such as spanning tree or maximal matching during the changes in the network and aim to use less update time as possible.

The maximal independent set (MIS) problem plays an important role in graph theory. By finding the MIS, one can approximately solve other basic problems such as vertex cover, matching, vertex coloring and edge coloring(see Linial\cite{linial1987distributive}).

The problem of maintaining an MIS in dynamic graph has a trivial solution of amortized running time $O(\Delta)$, where $\Delta$ is the maximum degree of vertices in the graph and can be as large as $m$, where $m$ is the number of dynamic changing edges. See details in \cite{2018arXiv180209709A}.
After Censor-Hillel \textit{et al.}\cite{censor2016optimal} left an open question of whether there is a non-trivial sequential algorithm to deal with fully dynamic MIS (their paper developed a randomized algorithm in distributed setting), the first solution was come up by Assadi \cite{2018arXiv180209709A} with $O(m^{3/4})$ amortized update time and it is deterministic. We presented a simple algorithm which is also deterministic and improved this result to $O(m^{2/3}\sqrt{\log m})$. Briefly speaking, we construct a good-MIS which defines one kind of MIS that every vertex with high degree has a bunch of neighbors in the MIS. If those neighbors run out during the adversary edge updates, we reconstruct the graph and do the same thing. Using this idea, we further study the randomized setting.

For randomized algorithm, we get a result of expected $O(\sqrt{m} \log^{1.5} m)$ amortized time against an oblivious adversary. Inspired by the paper about dynamic coloring \cite{bhattacharya2018dynamic}, when we add an edge between two vertices which are both in the MIS, the cost of removing one vertex from the MIS will be high. However, we can use a randomized algorithm to prevent the vertices with high degree from adding to the MIS. Our main idea is similar to the previous simple deterministic algorithm. Instead of finding a good-MIS using the greedy algorithm, we use a simple randomized algorithm. We find a random order on all the vertices and add them to the MIS one by one. We find a good property of this simple randomized algorithm. For a vertex $v$ with degree greater than $\Omega(\sqrt{m} \log^{1.5} m)$, the probability of this vertex joining in the MIS is $O(\frac{m \log^3 m}{d^2})$. When we add an edge between a vertex $u$ and $v$ and the vertex $u$ is removed from the MIS, the cost is $O(\deg(u))$. The probability that $u$ in the MIS is bounded by $O(\frac{m\log^3 m}{deg(u)^2})$. The expected cost is bound by $\min(\sqrt{m} \log^{1.5} m, \frac{m\log^3 m}{deg(u)})=O(\sqrt{m} \log ^{1.5}m)$. We use this randomized algorithm to construct an MIS in every $\lceil \sqrt{m}\rceil$ edge updates. After finding the MIS, we maintain the counter and update the MIS directly in the following steps.

%Things seem become easy when we get a sparse graph. But it turns out to be hard since the maximum degree can be as large as $O(m)$ even in low-arboricity graphs. To solve this problem, we purpose a deterministic algorithm which runs in $O(\sqrt{m\log m}\lambda^{\frac{3}{2}})$ amortized time, provided that the dynamic graph has arboricity at most $\lambda$ at all time. To achieve a $o(\sqrt{m})$ result is still a barrier even in a fully dynamic forest if we are required to show MIS explicitly,  as we explain the reason in the last section. There is still a huge gap between this problem and many other fully dynamic graph problems. For example,... achieves ......... From another view, we can say their "hierarchy" of locality is different. 

%{\color{red}{[Explain some hardness in arboricity, list some result about other dynamic problem e.g. bipartite matching that related to arboricity. briefly explain the gap.]}}

%We start with a $O(m^{2/3}\sqrt{\log m})$ deterministic algorithm in the section 2. This algorithm sets a framework for the following dynamic MIS algorithms. We describe a $O(\sqrt{m} \log^{1.5} m)$ randomized algorithm in the section 3. And a deterministic algorithm with $O(\sqrt{m\log m}\lambda^{\frac{3}{2}})$ amortized running time in the section 4. Finally, we discuss the hardness of this problem in section 5. {\color{red} {You should BB some words about hardness. }}

We start with a $O(m^{2/3}\sqrt{\log m})$ deterministic algorithm in the section \ref{section_alg_1}. This algorithm sets a framework for the following dynamic MIS algorithms. We describe a $O(\sqrt{m} \log^{1.5} m)$ randomized algorithm in the section \ref{section3}. Finally, we discuss the hardness of this problem in section \ref{hardness}.

%% file: determinstic.tex
\section{$O(m^{2/3}\sqrt{\log m})$ deterministic algorithm}\label{section_alg_1}
Recall that maximal independent set(MIS) of an undirected graph is a $maximal$ collection of vertices subject to the restriction that no pair of vertices in the collection are adjacent.
We focus on fully dynamic MIS which aims to maintain an MIS of a dynamic graph $G$, denoted by $\mathcal{M}=\mathcal{M}(G)$, subject to a sequence of edge insertions and deletions. Such an edge insertion or deletions is called an \textit{edge update}(or one \textit{round} for short). Our result is the following:

\begin{theorem}\label{1}
Starting from an empty graph on n fixed vertices, a maximal independent set can be maintained deterministically over any sequence of edge insertions and deletions in $O(m^{2/3}\sqrt{\log m})$ amortized update time, where m denotes the number of dynamic edges.
\end{theorem}

We will first introduce the following lemma to help proof Theorem $\ref{1}$.

\begin{lemma}\label{2}
Starting with any arbitrary graph on n vertices and $m$ edges, a maximal independent set $\mathcal{M}$ can be maintained deterministically over any sequence of $K \leq \lceil m^{1/3} \rceil $ edge insertions and deletions in \hbox{$O(K\cdot m^{2/3}\sqrt{\log m} + m)$} time.
\end{lemma}

We first show that this lemma implies Theorem \ref{1}.

\proof[Proof of Theorem \ref{1}]
\begin{comment}
Start from an empty graph, we can use brute force to maintain dynamic MIS until the number of dynamic edges reaches 100, denoted by $m_1=100$. Since $K_1=\lceil m_1^{1/3} \rceil\neq 0$, we can apply \hbox{lemma $\ref{2}$} for the next $K_1$ dynamic edges, and this time the number of dynamic edges reaches $m_2=100+K_1$. In fact, if we iteratively apply lemma \ref{2} by setting $m=m_t,~~K_t=\lceil m_t^{1/3} \rceil,~~m_{t+1}=m_t+K_t$, we will finally get an algorithm running in $O(m^{2/3}\sqrt{\log m})$ amortized update time where $m$ now denotes the number of total number of dynamic edges.$\hfill\square$
\end{comment}
For simplicity, we can say $m=1$ in the case of empty graph. We define a process as following: we started by counting the number of edges in the graph, denoted by $m_t$. Calculate $K_t=\lceil m_t^{1/3} \rceil$. And then apply the algorithm in lemma \ref{2} to deal with the following $K_t$ dynamic edges. If we iteratively do this process, we can get the final result. $\hfill\square$

We will focus on proofing lemma \ref{2} in the following paragraph, we first define a notation and come up with some lemmas.

\begin{defn}
$V_{high}=\{v|\text{deg}(v)\geq 10m^{2/3}\sqrt{\log m}\}$, where $deg(v)$ is the degree of v at the beginning. $V_{high}$ will stay unchanged over later edge updates.
\end{defn}

\begin{lemma}\label{3}
Given a graph with n vertices and $m$ edges, there is an algorithm which can construct a good-MIS $\mathcal{S}$ in $O(m)$. Good-MIS means that $\forall v\in V_{high}, |N(v)\cap \mathcal{S}|\geq m^{1/3}+1$, where $N(v)$ denotes the set of whole neighbors of $v$ excluding $v$.
\end{lemma}

\proof We start the proof by presenting the algorithm \ref{algorithm_det1}.

\begin{algorithm}
\caption{}\label{algorithm_det1}
\begin{algorithmic}
\State $\mathcal{S}\gets \emptyset$
\State $V_h\gets V_{high}$
\State $P\gets V\setminus V_{high}$
\State $cost(v)\gets 0,~n(v)\gets 0, ~\forall v\in V_{high}$
\While{$V_h\neq \emptyset$}
    \State $u\gets \arg\min_{u\in P} \frac{|N(u)|}{|N(u)\cap V_h|}$
    \State $\mathcal{S}\gets \mathcal{S}\cup \{u\}$
    \State $cost(v)\gets cost(v)+\frac{|N(u)|}{|N(u)\cap V_h|},~\forall v\in N(u)\cap V_h$
    \State $n(v)\gets n(v)+1,~\forall v\in N(u)\cap V_h$
    \State $V_h\gets V_h\setminus \{v\in V_h\mid n(v)=\lceil m^{1/3}\rceil +1\}$
    \State $P\gets P\setminus N(u)$
\EndWhile
\State $\mathcal{S}\gets \mathcal{S}\cup \{\text{MIS of graph induced by }P\}$
\end{algorithmic}
\end{algorithm}

Denote $h=|V_{high}|$ for simplicity. We will proof the correctness of this algorithm by showing that $\forall v\in V_{high},~|N(v)\cap P|\geq 5m^{2/3}\sqrt{\log m}$ for all time. By contradiction, if this property is broken at some time, we focus on the first time the property is broken. We list the vertices which are kicked out from $V_h$ as $v_1,...,v_k$ in their deleting order. Notice that the total number of vertices removed from $P$, denote this number $x$, is bounded by $\sum_{v\in V_{high}}cost(v)$. And $cost(v_i)\leq \frac{2m}{(h-i+1)\cdot(5m^{2/3}\sqrt{\log m})}(m^{1/3}+1)$ since all vertices in $V_h$ has at least $5m^{2/3}\sqrt{\log m}$ neighbors in $P$ at the time it been removed. Further more, $\sum_{u\in P}|N(u)|\leq 2m$. So 
\begin{equation}\label{equationXX}
\begin{aligned}
x&\leq \sum_{v\in V_{high}}cost(v)\leq \sum_{i=1}^k\frac{2m(m^{1/3}+1)}{(h-i+1)\cdot(5m^{2/3}\sqrt{\log m})}+(h-k)\frac{2m(m^{1/3}+1)}{(h-k)\cdot(5m^{2/3}\sqrt{\log m})}\\&\leq \frac{3}{5}m^{2/3}\sqrt{\log m}.
\end{aligned}
\end{equation}

But all vertices in $V_{high}$ have at least $10m^{2/3}\sqrt{\log m}$ neighbors at initial. So this algorithm is correct.

\begin{comment}
We can maintain $\frac{|N(u)|}{N(u)\cap V_h}$ for all $u$ using a heap. Since the total changes in $\frac{|N(u)|}{N(u)\cap V_h}$ only happen at most $O(m)$ times, the running time of this algorithm is $O(m\log m)$.
\end{comment}

The bottleneck of the implementation of the above algorithm is to query $\arg\min_{u\in P} \frac{|N(u)|}{|N(u)\cap V_h|}$ for at most $m^{2/3}$ times while the total changes in $\frac{|N(u)|}{N(u)\cap V_h}$ can happen as large as $O(m)$ times. To see why it can be done in $O(m)$ time, a crucial observation is that we can choose $u\in P$ in \hbox{2-approximation} rather than finding the minimum one; this would only impact $x$ twice larger than the original in (\ref{equationXX}). Another observation is that all numbers are fraction number and their denominator is at most $O(m^{1/3})$. So we can divide those numbers into $\log m$ classes, the $i$th class consist of all numbers with denominator between $2^{i-1}$ and $2^i-1$. For each class, we can maintain a 5-layer segment tree with branch size $m^{1/4}$ to find the minimum numerator. For each change happens, we can update the segment tree in constant time. And for each query, we can enumerate all classes to find a 2-approximation of the minimum in $O(m^{1/4}\log m)$ time. Thus we can construct a good-MIS in linear time. $\hfill\square$

\proof[Proof of Lemma \ref{2}]

We first come up with an algorithm and then analyze its running time. In the beginning, we construct a \emph{good-MIS} $\mathcal{M}$ using the algorithm mentioned in lemma \ref{3}, and we will maintain on this MIS $\mathcal{M}$. For every vertex $v$ in the graph, we maintain a counter \texttt{MIS-counter[v]} which counts the number of its neighbors in $\mathcal{M}$. We will update the counter and the MIS $\mathcal{M}$ in the following way.

When an edge $e_t=(u_t,v_t)$ is deleted from the graph, the only non trivial case is that exactly one of its endpoint, say $u_t$, is in $\mathcal{M}$. We can just decrease \texttt{MIS-counter[$v_t$]} by 1. If \texttt{MIS-counter[$v_t$]} becomes 0, we should add it into $\mathcal{M}$ and update its neighbor's counter.

Now we suppose that $e_t$ is inserted into the graph, the hard case occurs when both $u_t$ and $v_t$ are in $\mathcal{M}$, otherwise we can just update in $O(1)$ time. We should remove one of $u_t$ and $v_t$ from $\mathcal{M}$. Assume it is $u_t$. After removing $u_t$ from $\mathcal{M}$ and updating $u_t$'s neighbor's counter, to ensure the maximality of $\mathcal{M}$, we should enumerate $u_t$'s neighbor to check whether their \texttt{MIS-counter[.]}$=0$. If so, we should add this new vertex to $\mathcal{M}$ and update its neighbor's \texttt{MIS-counter[.]}. We should iteratively do this process until there is no vertex $v$ s.t. \texttt{MIS-counter[v]}$>$0.

\begin{obs}\label{4}
$\forall v\in V_{high}, |N(v)\cap \mathcal{M}|> 0$ during the whole $K$ rounds.
\end{obs}

\proof $\forall v\in V_{high}$, at most one neighbor of $v$ will be removed from $\mathcal{M}$ each round. Since they have at least $\lceil m^{1/3}\rceil+1$ neighbors in $\mathcal{M}$ at the beginning, the statement of observation is correct. $\hfill\square$

Let the potential function $\Phi=-\sum_v \texttt{MIS-counter[v]}$. $\Phi$ has a lower bound of $-2(m+K)$, and it is always $\leq 0$. In each round that either an edge is inserted or being deleted, $\Phi$ will increase by at most $O(m^{2/3}\sqrt{\log m})$ because there is at most one vertex being removed from $\mathcal{M}$ and this vertex has degree $\leq 10m^{2/3}\sqrt{\log m}$ due to Observation \ref{4}. Each time a vertex $v$ join $\mathcal{M}$, the algorithm should enumerate the whole neighbor of $v$ to update their counters and do some checks with running time $O(deg(v))$. It seems that the algorithm spends lots of time on this since there will be many vertices joining $\mathcal{M}$ in one iterative process. But each time a vertex $v$ joining $\mathcal{M}$ will cause $\Phi$ decreasing by $deg(v)$ since every neighbor's counter is increased by $1$. So the running time of this algorithm is bounded by the time of construction part plus the variation of $\Phi$, which is $O(m+ K\cdot m^{2/3}\sqrt{\log m})$. $\hfill\square$

%% file: random.tex
\section{$O(\sqrt{m}\log^{1.5}m)$ randomized algorithm}\label{section3}

%We present an algorithm for maintaining a maximal independent set in a graph under addition and deletion of edges. Our algorithm is randomized and it takes expected amortized $O(\sqrt{m} \log^2 m)$ time for each edge update against an oblivious adversary where $m$ is the number of edge updates in the graph.

%Our main idea is similar to the deterministic algorithm. Instead of finding a good-MIS using greedy algorithm, we shuffle all the vertices randomly and add them to the MIS one by one. We find a good property of this simple randomized algorithm. For a vertex $v$ with degree greater than $\sqrt{m} \log^2 m$, the probability of this vertex joining in the MIS is $O(\frac{m \log^4 m}{d^2})$. When we add an edge between a vertex $u$ and $v$ and the vertex $u$ is removed from the MIS, the cost is $O(deg(u))$. The probability that $u$ in the MIS is bounded by $O(\frac{m\log^4 m}{deg(u)^2})$. The expected cost is bound by $\min(\sqrt{m} \log^2 m, \frac{m\log^4 m}{deg(u)})=O(\sqrt{m} \log ^2m)$. We use this randomized algorithm to construct an MIS in every $\lceil \sqrt{m}\rceil$ edge updates. After we finding the MIS, we maintain the counter and update the MIS directly in the following steps.

%{\color{red}{[Yuhao: This part should be put in the introduction]}}

We describe a randomized algorithm for finding an MIS. We find a random order on the vertices. We take the vertex into the MIS one by one. Once a vertex is added into the MIS, we remove the vertex itself and all its neighbors. The algorithm itself is simple. But it's not easy to analyze this algorithm. In the paper \cite{censor2016optimal}, they also maintain the MIS constructed by this randomized algorithm. There is a theorem in this paper that there are $O(1)$ vertices changed their states after one edge update. But there is no result about the degree of vertices, which is important in the dynamic algorithm. We use a quite different argument to find the relation between the vertices with high degree and the probability that they join in the MIS. In the paper \cite{censor2016optimal}, the MIS after each update should be the same as one constructed by the greedy algorithm. So the change of one vertex may cause a ``chain reaction''. It will be harder to bound the expected degree of vertices that changed their states after one edge update. 

Instead of maintaining the MIS iteratively, we maintain the MIS in a similar manner with the previous algorithm. Once a vertex changed its state, it will only change the states of its neighbors. It is easier to analyze this algorithm. To make the MIS not too far from the MIS constructed by the randomized algorithm, we reconstruct the MIS in $O(\sqrt{m})$ rounds.

We first want to analyze the static randomized algorithm for MIS. For a vertex with high degrees, its neighbors will be removed by other vertices. The first lemma is about a bipartite graph. You can regard one part as the neighbors of one vertex with high degrees, the other part as other vertices.

\subsection{A static randomized algorithm for MIS}

\begin{lemma}\label{randlemma}
Given a bipartite graph $G=(A,B,E), |E| = m, |A|=d\geq 100\sqrt{m}\log^{1.5} m$. We find a random order $\sigma$ on the vertices, where $\sigma(u)$ is the position of vertex $u$ in the permutation. $$\Pr_{\sigma}[\forall u\in A, \sigma(u) \geq \min_{v\in N(u)} \sigma(v)]\leq \frac{1300 m\log ^3m}{d^2}.$$
\end{lemma}

\proof[Proof of Lemma \ref{randlemma}]

Let $L=\lceil\log \frac{d}{100 \log^2 m}\rceil, p=\lceil 6\log m\rceil$.

We partition the vertices in to $L+2$ sets. For all $i \in [0,L]$, $B_i=\{v\in B|deg(v) \in[2^i, 2^{i+1})\}$. $B_{+\infty} = \{v \in B | deg(v) \geq 2^{L+1}\}$. So $|B_i| \leq \frac{m}{2^i}, |B_{+\infty}| \leq \frac{100 m \log^2m}{d}$.

We say $u$ is controlled by $v$ if $v\in N(u)$ and $\sigma(v) < \sigma(u)$. 

Let $A'(\sigma)$ be the first $p$ vertices in the set $A$ in the random permutation $\sigma$.

$$\Pr_{\sigma}[\forall u\in A, \sigma(u) \geq \min_{v\in N(u)} \sigma(v)] \leq \Pr_{\sigma}[\forall u\in A'(\sigma), \sigma(u) \geq \min_{v\in N(u)} \sigma(v)]$$

So we only consider the first $p$ in the set $A$ instead of the whole set $A$. For simplicity, $A'$ denotes $A'(\sigma)$ in the following proof.

First we consider the case $$\min_{v \in B_{+\infty}} \sigma(v) < \max_{u \in A'} \sigma(u)$$

\begin{lemma}{\label{randlemmasc}}
Let $C$ be a vertex set disjoint with $A$. $\sigma_c$ is the permutation $\sigma$ restricted on $A$ and $C$. $$\Pr_{\sigma_c}[\min_{v \in C} \sigma_c(v)<\max_{u\in A'(\sigma_C)} \sigma_c(u)] \leq \frac{2p|C|}{d}.$$
\end{lemma}

\proof[Proof of Lemma \ref{randlemmasc}]
If $\min_{v \in C} \sigma(v) > \max_{u \in A'(\sigma_C)} \sigma(u)$, the first $p$ vertices are all the vertices in $A$. The probability that the first vertex is in the set $A$ in $\sigma_c$ is $\frac{d}{d+|C|}$. Then the probability that the second vertex in set $A$ is $\frac{d-1}{d+|C|-1}$.

So the probability that $\min_{v \in C} \sigma_c(v) > \max_{u \in A'} \sigma_c(u)$ is

%\begin{eqnarray*}
 %   &  & \prod_{i=0}^{p-1} \frac{d-i}{d+|C|-i} \\
  %  & = & \prod_{i=0}^{p-1} \left(1-\frac{|C|}{d+|C|-i} \right) \\
   % & \geq & 1 - \sum_{i=0}^{p-1} \frac{|C|}{d+|C|-i} \\
    %& \geq & 1 - \frac{p|C|}{d+|C|-p}\\
%    & \geq & 1 - \frac{2 p|C|}{d}.
%\end{eqnarray*}

\begin{equation*}
\prod_{i=0}^{p-1} \frac{d-i}{d+|C|-i} \geq 1 - \sum_{i=0}^{p-1} \frac{|C|}{d+|C|-i} \geq  1 - \frac{p|C|}{d+|C|-p} \geq  1 - \frac{2 p|C|}{d}
\end{equation*}

So

\begin{equation*}
    \Pr[\min_{v \in C} \sigma_C(v) < \max_{u \in A'(\sigma_C)} \sigma_C(u)] \leq \frac{2 p|C|}{d}.
\end{equation*} $\hfill\square$

Let $C$ in the lemma \ref{randlemmasc} be $B_{+\infty}$ and $\sigma_c$ be $\sigma_{+\infty}$. We can get $$\Pr[\min_{v \in B_{+\infty}} \sigma(v) < \max_{u \in A'} \sigma(u)] \leq \frac{2p|B_{+\infty}|}{d} \leq \frac{1299 m\log^3m}{d^2}.$$

Then we consider the case that $\min_{v \in B_{+\infty}} \sigma(v) > \max_{u \in A'} \sigma(u)$. In this case, no vertex in $B_{+\infty}$ appears very early in the permutation. So all the vertices in $A'$ is controlled by vertices in set $B_1, B_2, \dots, B_L$.

The vertices in $A'$ can only be controlled by the vertices appear earlier than the last vertices in $A'$. We want to show that with high probability, there will not be too many vertices in $B_i$ that appear earlier than the last vertices in $A'$.

Let $B'_i=\{v \in B_i|\sigma(v) < \max_{u \in A'(\sigma)} \sigma(u)\}$.

\begin{lemma} \label{randlemma2}
The probability that $|B'_i|\geq \frac{8(d+|B_i|)\log^2 m}{d}$ is less than $\frac{1}{m^3}$. 
\end{lemma}

\proof[Proof of Lemma \ref{randlemma2}]

Let $b=|B_i|, k=\lceil \frac{8(d+b)\log^2 m}{d} \rceil$. 

If there are exact $t$ vertices in the set $B_i$ appear before the $p$-th vertex in the set $A$, the probability is $\frac{\binom{p-1+t}{t}\binom{d-p+b-t}{b-t}}{\binom{d+b}{d}}$.

So 

\begin{eqnarray*}
& & \Pr[|B'_i| \geq k] \\
& = & \sum_{t=k}^{b} \frac{\binom{p-1+t}{t}\binom{d-p+b-t}{b-t}}{\binom{d+b}{d}} \\
& \leq & \frac{\binom{p-1+b}{b}}{\binom{d+b}{d}} \sum_{t=k}^{b} \binom{d-p+b-t}{b-t} \\
& = & \frac{\binom{p-1+b}{b}\binom{d-p+b-k+1}{d-p+1}}{\binom{d+b}{d}} \\
& = & \frac{(b+p-1)!(d-p+b-k+1)!d!b!}{b!(p-1)!(d-p+1)!(b-k)!(d+b)!} \\ 
& = & \binom{d}{p-1} \frac{(b-k+1)(b-k+2)\cdots(b-k+1+d-p)}{(b+p)(b+p+1)\cdots (d+b)} \\
& \leq & \binom{d}{p-1} \left(\frac{d-p+b-k+1}{d+b}\right)^{d-p+1} \\
& = & \binom{d}{p-1} \left(1-\frac{p+k-1}{d+b}\right)^{d-p+1} \\
& \leq & d^p e^{-\frac{d(p+k-1)}{d+b}} \\
& \leq & e^{p\log d -\frac{dk}{d+b}} \\
& \leq & e^{p \log d - 8 \log^2 m} \\
& \leq & m^{-\log m}.
\end{eqnarray*}

We finish the proof of lemma $\ref{randlemma2}$. By the union bound, with probability $1-\frac{1}{m^2}$, $|B'_i| \leq \frac{8(d+|B_i|)\log^2 m}{d}$ for all the layer. $\hfill\square$

So the total number of vertices in $B'_1, B'_2, \dots, B'_L$ can control is less than 

\begin{eqnarray*}
& & \sum_{i=0}^L |B'_i|2^{i+1} \\
& \leq & \sum_{i=0}^L 8\left(1+\frac{|B_i|}{d} \right)2^{i+1} \log^2 m \\
& \leq & \sum_{i=0}^L 8\left(1+\frac{m}{2^i d} \right)2^{i+1} \log^2 m \\
& \leq & 16\frac{m \log^2 m(L+1)}{d} + 8 \sum_{i=0}^L 2^{i+1} \log^2 m \\
& \leq & \frac{16m\log ^3 m}{d}+\frac{32d}{100} \\
& \leq & \frac{d}{2}.
\end{eqnarray*}

It means that with probability $1-\frac{1}{m^2}$, the vertices in $B'_1, B'_2, \dots, B'_L$ can only control no more than half vertices. The choices of $A'$ and $B'_i$ are independent. So for every vertex in $A'$, the probability that it is controlled is no more than $\frac{1}{2}$. The probability that all the vertices in $A'$ are controlled is no more than $\frac{1}{2^p} \leq \frac{1}{m^3}$.

Combining these two cases, we finish the proof of lemma \ref{randlemma}. $\hfill\square$

\begin{cor}\label{randcor1}
Given a bipartite graph $G=(A,B,E), |E| = m, |A|=d\geq 100\sqrt{m}\log^{1.5} m$. We find a random order $\sigma$ on the vertices, where $\sigma(u)$ is the position of vertex $u$ in the permutation. The probability that all the first $\lceil6\log m\rceil$ vertices in $A$ are controlled is no more than $\frac{1300 m\log ^3m}{d^2}$.
\end{cor}

\begin{theorem}\label{randalgothm}
After running the randomized algorithm, for a vertex $w$ with degree greater than $\sqrt{m} \log^{1.5} m$, the probability that $w$ is added into the MIS is no more than $\frac{1301m\log ^3 m}{d^2}$.
\end{theorem}

\proof[Proof of Theorem \ref{randalgothm}]

Let $d$ be the degree of $w$. Let $A=N(w), B=V\backslash (A\bigcup \{w\})$.

There are two cases that $w$ will be added to the MIS. The first case is that $u$ appears very early and is added to the MIS. The second case is that nearly all the vertices in $A$ are removed. So $w$ can't be removed by its neighbor. 

We consider the first case. We can apply the lemma \ref{randlemmasc} to bound the probability that $w$ appears before the first $\lceil 6\log m\rceil$ vertices in $A$. Let $C$ in lemma \ref{randlemmasc} be $\{w\}$, so the probability is less than $\frac{13m\log m}{d^2}$. 

Then we consider that $w$ appears later than the first $\lceil 6\log m\rceil$ vertices. If all the first $\lceil 6\log m\rceil$ vertices are removed without joining the MIS, these vertices can only be removed by its neighbor in set $B$. Applying the corollary \ref{randcor1}, the probability is no more than $\frac{1300 m\log ^3m}{d^2}$.

Combining these two cases, we finish the proof of the theorem \ref{randalgothm}. The probability that vertex $w$ appears before the first $\lceil 6\log m\rceil$ vertices in $A$ or none of the first $\lceil 6\log m\rceil$ vertices are in the MIS is no more than $\frac{1301m\log^3 m}{d^2}$. $\hfill\square$

\subsection{A randomized algorithm for fully dynamic MIS}

\begin{defn}
$V_{high}=\{v|deg(v)\geq 200\sqrt{m}\log^{1.5} m\}$, where $deg(v)$ is the degree of $v$ after the construction of the MIS.
\end{defn}

We use a similar algorithm in the previous part. In the beginning, we use the randomized algorithm to construct an MIS $\mathcal{M}$. For every vertex $v$ in the graph, we maintain a counter \texttt{MIS-counter[v]} which counts the number of its neighbors in $\mathcal{M}$. We will update the counter and the MIS $\mathcal{M}$ in the same way. When an edge is deleted, we add the vertex into the MIS if possible. When an edge is added between two vertices in the MIS, we remove one vertex arbitrarily.

We reconstruct the MIS in every $\lceil\sqrt{m}\rceil$ rounds.

The challenge is analyzing the running time. First, we will give a dynamic version of the theorem $\ref{randalgothm}$. 

\begin{theorem}\label{randdynprobthm}
For a vertex $w \in V_{high}$, the probability that $w$ is added into the MIS after $K\leq \sqrt{m}$ edge updates is no more than $\frac{1302m\log ^3 m}{deg(w)^2}$.
\end{theorem}

\proof[Proof of Theorem \ref{randdynprobthm}]

Let $\mathcal{M}_0$ be the MIS constructed by the randomized algorithm. And let $\sigma$ be the random order to construct the MIS.

First, we can see that every edge updates will cause no more than one vertex to be removed from the MIS. We will only remove a vertex from the MIS if we add an edge between two vertices in the MIS.

For a vertex $w \in V_{high}$, let $A$ be its neighbors after construction of the MIS. $B=V\backslash (A\bigcup \{w\})$.  Once a vertex $u$ in $A$ is removed from the MIS, or the edge $(w,u)$ is deleted, we removed the vertex $u$ from the set $A$ permanently. The vertex set $A$ will become smaller and smaller. However, we can only remove at most one vertex from $A$ in each round. $|A|$ will be no less than $200\sqrt{m}\log^{1.5} m -\sqrt{m}\geq 100\sqrt{m} \log^{1.5} m$. If $w$ appears later than the first $\lceil 6\log m\rceil$ vertices in the $\sigma$ and one vertex among the first $\lceil 6\log m\rceil$ vertices are in the $\mathcal{M}_0$, $w$ can't be added into the MIS. So the probability that $w$ is added into the MIS is less than $\frac{1301 m\log ^3 m}{|A|^2} \leq \frac{1302 m \log^3 m}{deg(w)^2}$. $\hfill\square$

\begin{theorem}\label{randdyntimethm}
The expected total cost in $K\leq \sqrt{m}$ edge updates after the construction of the MIS is $O(m\log^{1.5}{m})$.
\end{theorem}

\proof[Proof of Theorem \ref{randdyntimethm}]

Let the potential function $\Phi=-\sum_v \texttt{MIS-counter[v]}$. $\Phi$ is always between $-m$ and $0$. Once one vertex $u$ is added into or removed from the MIS, the cost of updating $\texttt{MIS-counter}$ is $O(deg(u))$, $\Phi$ is increased or decreased by $deg(u)$ respectively. We only need to care about the total increasing amount of $\Phi$. The total decreasing amount of $\Phi$ is at most the total increasing amount plus $m$. $\Phi$ will be increased only if we added an edge between vertices $u$ and $v$.

If one vertex of $u$ and $v$ is not in $V_{high}$. We can choose to remove the vertex that is not in the $V_{high}$. the decreasing amount is no more than $O(\sqrt{m} \log^{1.5} m)$. If both vertices are in $V_{high}$, the probability that $u$ in the MIS is no more than $\frac{1302 m \log^3 m}{deg(u)^2}$. The expected increasing amount is no more than $\frac{1302 m \log^3 m}{deg(u)} \leq 14\sqrt{m}\log^{1.5} m$ in one edge deletion. So the expected total increasing amount is $O(m\log^{1.5} m)$ in $\sqrt{m}$ edge updates. $\hfill\square$

\begin{cor}\label{randcomplexity}
Starting from an empty graph on $n$ fixed vertices, a maximal independent set can be maintained randomized against an oblivious adversary over any sequence of edge insertions and deletions in $O(\sqrt{m}\log^{1.5} m)$ expected amortized update time, where $m$ denotes the number of dynamic edges.
\end{cor}

%% file: hardness.tex
\section{Discussion on Hardness}\label{hardness}
\subsection{Is MIS-counter necessary?}
%As the above showed that all of our algorithms is using \texttt{MIS-counter}, it is indeed a convenient way to check whether a vertex should join $\mathcal{M}$ after some modifications. But this will give us an amortized $\Omega(\sqrt{m})$ lower bound on maintaining the exact \texttt{MIS-counter} no matter in the oblivious adversary case or low arboricity case. The main idea is to do edge update between two vertices in $\mathcal{M}$ with degree $\Omega(\sqrt{m})$ incessantly.
As the above showed that all of our algorithms is using \texttt{MIS-counter}, it is indeed a convenient way to check whether a vertex should join $\mathcal{M}$ after some modifications. But this will give us an amortized $\Omega(\sqrt{m})$ lower bound on maintaining the exact \texttt{MIS-counter} in the oblivious adversary case. The main idea is to do edge update between two vertices in $\mathcal{M}$ with degree $\Omega(\sqrt{m})$ incessantly. 

We can construct a complete bipartite graph $K_{\sqrt{m},\sqrt{m}}$ with each side $\sqrt{m}$ vertices. There are only two kinds of MIS since, by definition of MIS, it either contains the whole left side vertices or the whole right side vertices. And we call one round by adding an edge between two arbitrary vertices in the left side and another edge linking two arbitrary vertices in the right side and delete both of these two edges. In one such round, there must be one vertex removed from $\mathcal{M}$. Thus we must take  $\Omega(\sqrt{m})$ to maintain the \texttt{MIS-counter} per round. If we do $m$ rounds, we can get the $\Omega(\sqrt{m})$ lower bound. 

%For low arboricity case, we can even consider a dynamic forest which has arboricity $\lambda=1$. We can construct a three-layer rooted tree. The first layer only contains one vertex which is called root, and it has $\sqrt{m}$ child in the second layer. For each child of the root, it also has $\sqrt{m}$ child in the third layer. Thus there must be at least one vertex in the first two layers that are in MIS. If we do a copy of this tree, and add an edge between two such vertices and delete this edge for many times, we can also get the $\Omega(\sqrt{m})$ lower bound for this case.

So we must abandon \texttt{MIS-counter} or recording partial information about it to break through the $o(\sqrt{m})$ barrier. But if we do so, it becomes hard to maintain the maximality of $\mathcal{M}$ which requires you to answer whether one vertex has any neighbors in the MIS. \textit{OMv} hypothesis \cite{henzinger2015unifying} is a powerful hypothesis for providing a conditional lower bound for dynamic graphs. As we showed an imperfect reduction to the \textit{OMv} problem by the condition of forcing one vertex in $\mathcal{M}$ permanently.

\subsection{Imperfect Reduction to \textit{OMv} Problem}

We use the statement about \textit{OMv} hypothesis in the article \cite{williamssome}.

\begin{hypo}
Every (randomized) algorithm that can process a given $n \times n$ Boolean matrix $A$, and then in an online way can compute the products $Av_i$ for any $n$ vectors $v_1, \dots, v_n$, must take total time $n^{3-o(1)}$.
\end{hypo}

\begin{lemma}
If one algorithm $\mathcal{A}$ will force one vertex $w$ to be in MIS permanently, then $A$ cannot solve the fully dynamic maximal independent set problem in $O(m^{1/2-\epsilon})$ amortized time for any constant $\epsilon$, unless the OMv hypothesis is false. Where $m$ is the number of edge updates.
\end{lemma}

\proof Here we will show a reduction to \textit{OMv} problem. For one instance of Online Boolean Matrix Multiplication with $n \times n$ Boolean matrix $M$ and online sequence of vectors $v_1,...,v_n\in\{0,1\}^n$. We first construct a bipartite graph with vertices $\{1,...,n\}$ in left side and $\{1',...,n'\}$ in right side. $i$ and $j'$ have an edge if and only if $M_{i,j}=1$. When we receive $v_t$ as our query, we reconstruct the incident edges of $w$ to be every vertex $j'$ that $v_t(j)=0$ and all vertices in the left side. Now there is only one kind of MIS which is the set of $j'$ s.t. $v_t(j')=1$ and the vertex $w$ since the algorithm $\mathcal{A}$ will force $w$ to be in MIS. We denote this MIS by $X$. Then we will answer the result of $Mv_t$ line by line. The $i$th line correspond to vertex $i$. Now we delete the edge $(w,i)$. If $Mv_t(i)=0$, which means $N(i)\cap X=\emptyset$, $i$ must join the MIS. On the other hand, if $Mv_t(i)=1$, then there are two situations. Either $i$ do not join the MIS, or $i$ join the MIS and some vertices in $X$ get removed. If we trace the difference in MIS, we can answer $Mv_t(i)$ in constant time. Then we delete the edeg $(w,i)$ and start the next line. As a result, we used $O(n^2)$ edge updates and successfully solved this instance. $\hfill\square$